\documentclass[a4paper,11pt]{article}
\usepackage{jheppub} % for details on the use of the package, please
\pdfoutput=1 % if your are submitting a pdflatex (i.e. if you have
\usepackage[T1]{fontenc} % if needed
\usepackage{hyperref}
\usepackage{textcomp}

\usepackage[dvipsnames,table,xcdraw]{xcolor}

\title{\boldmath On the importance of factorization for fast binned likelihood inference}

\author{C.Jes\'us-Valls}

\affiliation{Center for Data-Driven Discovery, Kavli IPMU (WPI), UTIAS, The University of Tokyo, Kashiwa, Chiba 277-8583, Japan}

\emailAdd{cesar.jesus-valls@ipmu.jp}

\pdfoutput=1 % if your are submitting a pdflatex (i.e. if you have
\usepackage[dvipsnames,table,xcdraw]{xcolor}
\usepackage{graphicx}

\begin{document}

\abstract{Likelihood-based inference, central in modern particle physics data analysis requires the extensive evaluation of a likelihood function that depends on set of parameters defined by the statistical model under consideration. If an analytical expression for the likelihood can be defined from first principles the procedure is computationally straightforward. However, most experiments require approximating the likelihood numerically using large statistical samples of synthetic events generated using Monte Carlo methods. As a result, the likelihood consists of a comparison of the expected versus the observed event rates in a collection of histogram bins, defining binned likelihood functions. When this occurs, evaluating the likelihood function involves, on each occasion, recalculating the prediction in those bins, increasing the computational load of these analysis drastically. In this text, I highlight the importance of identifying which are the unique event configurations in the binned likelihood definition and I provide an exact formula to update the event rate predictions utilizing the minimum number of necessary calculations by means of factorization. The aim of the discussion is to decrease the computational load of widespread high-energy physics analyses, leading to substantial speed improvements and reduced carbon footprints.}

%The discussion aims at speeding up computationally heavy calculations in high energy physics and its associated carbon footprint.}
%, capable of remarkable speed ups that depend on the factorizability of the system.}

\maketitle

\section{Introduction}

% general intro
Likelihood-based inference is a mainstream statistical framework for the analysis of real and simulated data in contemporary high energy physics (HEP)~\cite{Cowan:1998ji}. The foundational assumption is that the data observables, represented by $\vec{x}$, stem from a known probability distribution, $f_{\vec{\theta}}$, determined by a set of parameters, $\vec{\theta}$. Bayesian and frequentist methods are used to draw conclusions through the study of the likelihood function, $\mathcal{L}$, which defines the probability of observing the collected data for any given value of $\vec{\theta}$, treating the data as fixed. To achieve this, one needs to calculate the likelihood of $\vec{\theta}$ given the observed data $\vec{x}$, conventionally expressed as $\mathcal{L}(\vec{\theta} \,|\, \vec{x})$ = $f_{\vec{\theta}}$. In some occasions, the system under study is simple enough as to allow for $f_{\vec{\theta}}$ to be evaluated directly from the analytical expressions characterizing it. In most cases, however, the accurate consideration of nontrivial theoretical quantities and experimental settings results in the necessity of using Monte Carlo (MC) methods. MC predictions are built upon finite samples of discrete data, resulting in three major consequences: 1) the extensive use of binned likelihoods, 2) the necessity of event reweighting and 3) and a largely increased computational load to evaluate the likelihood function. These three aspects play a central role in many studies in HEP, each imposing limitations that demarcate what physics studies are possible in practice. However, whereas the two first points are mainstream knowledge in most experiments and are often addressed in the scientific literature, e.g. Refs.~\cite{Cowan:2010js,Buckley:2011ms, Campbell:2022qmc}, the latter has been left out of the main discussion. In this article, I comment on the importance of using factorization in likelihood-based tests involving event reweighting and I illustrate generically how to implement such factorization with the goal of increasing the speed and reducing the carbon footprint~\cite{allen2022huge} of computationally heavy and widespread analysis in HEP.

%reducing the computation times and associated carbon footprint~\cite{allen2022huge}.

%of widespread HEP analysis.

\section{About the use of binned likelihoods}
\label{sec:sec_BL}

To commence, let's review why binned likelihood analysis are ubiquitous HEP analysis. In HEP, information is typically grouped in events, $\vec{n}$, each resulting from the observation of a fundamental interaction and consisting of the collection of the information measured by a detector. Observables $\vec{x}$ are built performing operations on the events' information, namely $\vec{x}=\mathcal{Q}(\vec{n})$. Given a probability distribution for those observables, $f_{\vec{\theta}} (\vec{x})$, determined by a set of parameters, $\vec{\theta}$, then it is possible to do inference on those parameters by studying the likelihood function $\mathcal{L}$, defined as:
\begin{equation}
    \label{eq:LL_def}
    \mathcal{L}(\vec{\theta} \,|\, \vec{x}) = f_{\vec{\theta}} (\vec{x}) = \prod_i f_{\vec{\theta}} (x_i).
\end{equation}
In general, analysis often consist of looking for the value of $\vec{\theta}$ that maximizes the above function, determining how likely are other values of $\vec{\theta}$ given the observed data, and studying the overall suitability of $f_{\vec{\theta}} (\vec{x})$ to be a faithful model representation of the data. When the probability of observing any given $\vec{x}$ can be calculated, studying Eq.~\ref{eq:LL_def} is computationally straightforward. However, in HEP experiments calculating such probability is oftentimes analytically impractical and Monte Carlo (MC) methods are used to sample synthetic events according to the underlying model probability density function. As a result, evaluating the likelihood infinitesimally is no longer possible and to overcome this limitation the likelihood is build through the comparison of the expected and observed distributions in finite regions in observable space, corresponding to the bin content of histograms of arbitrary dimensionality. Consequently, HEP experiments very often base their physics studies on binned likelihood analyses using:
\begin{equation}
\label{eq:ProdPoissonLL}
    \mathcal{L}(\vec{\theta} \,|\, \vec{N}_{\text{obs}}) = \text{P}(\vec{N}_{\text{obs}}  \,|\,  \vec{\lambda}(\vec{\theta})) = \prod_i\text{P}(N^i_{\text{obs}}  \,|\,  \lambda^i(\vec{\theta})),
\end{equation}
where $\vec{N}_{\text{obs}}$ and $\vec{\lambda}(\vec{\theta})$ denote respectively the collection of observed and expected events in all the observables bins $i$ and where P is a discrete probability distribution. If the counts in every bin are independent, as it is normally the case in HEP, then $P$ is known to be characterized, asymptotically, by the Poisson distribution. Since the sample of synthetic data is finite, however, the error of the MC sample plays a role that can be relevant in some conditions. Consequently, modified Poisson likelihoods that include these corrections are available~\cite{Barlow:1993dm, Arguelles:2019izp}.\\
Remarkably, regardless of the likelihood choice, in Eq.~\ref{eq:ProdPoissonLL} $\vec{N}_{\text{obs}}$ is treated as fixed and therefore the complexity of the calculation depends entirely on the difficulty of evaluating $\vec{\lambda}(\vec{\theta})$.

%Consequently, this is the central This is hence the central point of discussion in this article.\\

\section{Event reweighting}
The calculation of $\vec{\lambda}(\vec{\theta})$ using Monte Carlo techniques is among the most demanding computational tasks in HEP, often requiring in large experiments the use of hundreds or thousands of CPU hours to realize a single simulation, and resulting in huge computational demands every year~\cite{belyaev2017high}. Therefore, re-calculating $\vec{\lambda}(\vec{\theta})$ from scratch for every value of $\vec{\theta}$ in order to analyze $\mathcal{L}$ is computationally prohibitive. Instead, modern experiments use event re-weighting. To explain it, let's introduce some definitions that will become key later.\\
Consider the nominal expected event rate in the $i$-th bin as described by:
\begin{equation}
\label{eq:NaiveEvtSum}
\lambda_i(\vec{\theta}_{\text{nom}}) = \sum^{\text{events}}_j w^{ij}_{\text{nom}}
%\sum^{\text{i-events}}_j w^j_{\text{nom}}
%\lambda_i(\vec{\theta}_{\text{nom}})\times g_i(\vec{\theta}).
\end{equation}
where $w^{ij}_{\text{nom}}$ is the nominal event weight for the $j$-th event in the $i$-th bin. As any event can only be present in one bin:
\begin{equation}
  w^{ij}_{\text{nom}} =
    \begin{cases}
      0 & \text{if the $j$-th event is not present in the i-$th$ bin.}\\
      w^{k}_{\text{nom}} & \text{if the $k$-th event belongs to the $i$-th bin.}
    \end{cases}       
\end{equation}
From this definition, we can simplify equation~\ref{eq:NaiveEvtSum} into:
\begin{equation}
\label{eq:SmartEvtSum}
\lambda_i(\vec{\theta}_{\text{nom}}) = \sum^{\text{i-events}}_k w^{k}_{\text{nom}}
\end{equation}
where "i-events" indicates that the sum happens exclusively over the indices of those events in the $i$-th bin.\\
Let's consider now the effect of choosing a value of $\vec{\theta}$ other than nominal. Instead of generating new MC samples, the solution consists in calculating event-by-event response functions $\vec{g}$ that re-weight the contribution of each event as a function $\vec{\theta}$ with respect to their nominal value:
\begin{equation}
\label{eq:SmartEvtRwg}
\lambda_i(\vec{\theta}) = \sum^{\text{i-events}}_k (w^{k}_{\text{nom}} \times g_k(\vec{\theta}) ).
\end{equation}
Notice that, by definition, $g_k(\vec{\theta}_{\text{nom}})=1$, which allows to recover Eq.~\ref{eq:SmartEvtSum} from Eq.~\ref{eq:SmartEvtRwg}. The reweight functions are connected to the fit parameters through functions that act on subsets of disjoint parameters, associated to parameter indices denoted by $\alpha$. Namely:
\begin{equation}
\label{eq:RspF}
g_k(\vec{\theta}) = \prod_\alpha^{\text{all disjoint $\alpha$}} h^\alpha_{k}(\theta_\alpha).
\end{equation}
where
% The relation between $g_k$ and $\vec{\theta}$ is:
% \begin{equation}
% \label{eq:RspF}
% g_k(\vec{\theta}) = \prod_\alpha^{\text{all disjoint $\alpha$}} h^\alpha_{k}(\theta_\alpha).
% \end{equation}
% Thus, $g_k(\vec{\theta})$ is nothing but the product of the weights that result from evaluating an analytic function $h^\alpha_{k}$ acting on a disjoint subset of parameters characterized by an index $\alpha$. This definition is crucial for the later discussion. Before focusing on the clarification of its meaning, however, let's first note that:
\begin{equation}
    h^\alpha_{k}(\theta_\alpha) =
    \begin{cases}
      1 & \text{if the $k$-th event is unaffected by $\theta_\alpha$.}\\
      w^\alpha_{k} & \text{if the $k$-th event is affected by $\theta_\alpha$.}
    \end{cases}       
\end{equation}
With this in mind, Eq.~\ref{eq:RspF} reads:
\begin{equation}
\label{eq:RspFSmart}
g_k(\vec{\theta}) = \prod_\alpha^{\text{k-relevant disjoint $\alpha$}} h^\alpha_{k}(\theta_\alpha).
\end{equation}
where "k-relevant" indicates that the product happens exclusively for those $\alpha$ indices relevant for the $k$-th event.\\
Now that the notation has been introduced, let's clarify the meaning of the former equations considering a set of examples below.

\subsection*{Fully disjoint example with one parameter}
Let's consider the simplest possible physical model where we have a single parameter. Then, $\alpha$ is trivially disjoint, and in Eq.~\ref{eq:RspFSmart} $\alpha$ is just an integer representing the $0$-th element of the 1-dimensional parameter vector $\vec{\theta}$.\\
Imagine that the parameter under consideration, $\theta_0$, accounts for the detector mass. In this case, all events are affected by this parameter:
\begin{equation}
\label{eq:SmartEvtRwg}
\lambda_i(\vec{\theta}) = \sum^{\text{i-events}}_k (w^{k}_{\text{nom}} \times h^0_{k}(\theta_0)),
\end{equation}
and, $h^0_{k}(\theta_0)$ is equal for all the events. Therefore, if we define this identical function by $h^0$, follows:
 \begin{equation}
\label{eq:SmartEvtRwgFactExample}
\lambda_i(\vec{\theta}) =  \sum^{\text{i-events}}_k (w^{k}_{\text{nom}} \times h^0_{k}(\theta_0)) = h^0(\theta_0) \times \sum^{\text{i-events}}_k w^{k}_{\text{nom}} =  h^0(\theta_0)\times \lambda_i(\vec{\theta}_{\text{nom}}).
\end{equation}
In this example, Eq.~\ref{eq:SmartEvtRwgFactExample} shows that full factorization is possible. Since $\theta_0$ accounts for the detector mass, $\theta_0$ should play the role of a normalization parameter. Namely, $\theta_0$=1 means that the detector mass is nominal, and $\theta_0=1.2$ implies an increase of 20\% with respect to that nominal value. More in general, $h^0(\theta_0) = \theta_0$. Thus, in Eq.~\ref{eq:SmartEvtRwgFactExample}, we observe that the result aligns with the logic: if the detector mass increases by 20\%, the event rate in all bins goes up by 20\%.

\subsection*{Fully disjoint example with two parameters}
Let's consider a more generic case where we have two parameters, $\theta_0$ and $\theta_1$. Let $\theta_0$ be the same parameter as in the previous example and we introduce a new normalization parameter, $\theta_1$, representing the cross section of a specific interaction channel and that affects only those events generated through it. In this case, Eq.~\ref{eq:RspFSmart} reads:
\begin{equation}
\label{eq:RspFSmartExample2}
g_k(\vec{\theta}) = \prod_\alpha^{\{0,1\} } h^\alpha_{k}(\theta_\alpha) =
    \begin{cases}
      h^0(\theta_0) \times h^1 (\theta_1) & \text{if the $k$-th event is associated to $\theta_1$.}\\
      h^0(\theta_0) & \text{otherwise.}
    \end{cases}
\end{equation}
Where we have used the fact that all $h^0_k$ and $h^1_k$ functions are identical regardless of $k$ and their value is represented by $h^0$ and $h^1$ respectively. As in the previous example, the equivalence of these functions would allow to factorize the calculation. Before focusing on this topic on the next section, let's conclude the explanation with another example.

\subsection*{General example with joint and disjoint parameters}
Finally, we focus on the general case. Consider that, in addition to the former parameters $\theta_0$ and $\theta_1$, we account in our model for the role of two flavor neutrino oscillations. To do so, we include two new parameters $\theta_2$ and $\theta_3$ playing the role of the physics parameters $\phi$ and $\Delta m^2$ respectively. It is important to recall that the physics equation governing this process is:
\begin{equation}
\label{eq:twoNuOsc}
f^{2\nu}_{\text{osc}}(\phi, \Delta m^2_{23}, E_\nu, L) =  \sin^2{2\phi}\sin^2\left ( 1.27\frac{\Delta m^2 L}{E_\nu} \right).
\end{equation}
Thus, this new example has two major novelties. First, in contrast with $\theta_0$ and $\theta_1$ that can be considered independently, $\theta_2$ and $\theta_3$ are intertwined. Therefore, it does not make sense for $\alpha$ to be either 2 or 3, rather, we define this index as a joint index that we express as [2,3]. Secondly, Eq.~\ref{eq:twoNuOsc} is, additionally a function of the length $L$ from the neutrino production to the detection point and its energy $E_\nu$. Whereas in most practical cases $L_\text{{fixed}}$ is a constant for all the events that depends on the experimental settings, the neutrino energy is rarely monochromatic such that $E_\nu$ needs to be considered on an event-by-event basis. Consequently, 
\begin{equation}
h^{[2,3]}_k(\theta_2,\theta_3)=f^{2\nu}_{\text{osc}}(\theta_2, \theta_3, E^k_\nu, L_\text{{fixed}})
\end{equation}
is different for every $k$, alike $h^0_k$ and $h^1_k$ that are equal for all events as discussed in the examples above.\\
Following this considerations, we have that Eq.~\ref{eq:RspFSmartExample2} becomes:
\begin{equation}
\label{eq:RspFSmartGenExample}
    g_k(\vec{\theta}) = \prod_\alpha^{\{0,1,[2,3]\} } h^\alpha_{k}(\theta_\alpha) =
    \begin{cases}
      h^0(\theta_0) \times h^1 (\theta_1) \times h^{[2,3]}_{k}(\theta_2, \theta_3) & \text{case 1}\\
      h^0(\theta_0) \times h^1 (\theta_1) &  \text{case 2}\\
      h^0(\theta_0) \times h^{[2,3]}_{k}(\theta_2, \theta_3) & \text{case 3}\\
      h^0(\theta_0) & \text{otherwise.}
    \end{cases}
\end{equation}
Where
\begin{itemize}
    \item Case 1: The $k$-th event is affected by all weights.
    \item Case 2: The $k$-th event is affected by all but oscillation weights. That would be the case for events produced by neutral currents.
    \item Case 2: The $k$-th event is affected by all $\theta_1$-related weights, i.e. those not generated through the interaction channel associated to $\theta_1$.
\end{itemize}
An important observation is that, in this example two new effects converge: the existence of joint indices "[2,3]" and the necessity of using event-by-event response functions imposed by its dependence with  $E^k_\nu$. This is a specific characteristic of this example, but not an universal condition. In other words, it is entirely possible to encounter scenarios where a disjoint parameter necessitates unique event-by-event response functions and, conversely, to have models with joint indices with identical response functions for all events.

\section{Event factorization}
Now that the use of event reweighting in the calculation of the binned likelihood has been formalized and illustrated by means of examples, we can turn our attention to the computational aspect of the problem.\\
Firstly, note that in the generic Eq.~\ref{eq:SmartEvtSum}, one iterates over all of the event is one bin. Thus, to update the value of all bins, one needs to iterate over the full set of events. Considering now Eq.~\ref{eq:RspF}, for each event is necessary to iterate over all $\alpha$ relevant for that event. In the worse computational case, every $\alpha$ is relevant for each event. This leads to an upper bound of the number of calculations $N^{\text{max}}_c$ to update the event rate:
\begin{equation}
N^{\text{max}}_c = N_{\text{sim}} \times N_{\text{alpha}},
%\text{Dim($\vec{n}_{\text{sim}}$)} \times \text{Dim()}
\end{equation}
Namely, the maximum number of operations corresponds to the product of the number of all simulated events $N_{\text{sim}}$ with the number of unique $\alpha$ configurations $N_{\alpha}$.\\
Secondly, let's look at Eq.~\ref{eq:SmartEvtRwgFactExample}. In the extreme case where there is only parameter, and its associated reweighting function acts identically on all events, we can achieve full factorization, such that to update the event rate in every bin, one needs to perform one operation. This settles the lower bound of the number of calculations $N^{\text{min}}_c$, that consist of one operation per bin:
\begin{equation}
N^{\text{min}}_c = N_{\text{bins}}.
\end{equation}
Therefore, we conclude that in general the number of calculation, $N_c$, necessary to update $\vec{\lambda}(\vec{\theta})$ must be:
\begin{equation}
\label{eq:NcSpan}
N_{\text{bins}} \leq N_c \leq N_{\text{sim}} \times N_{\text{alpha}}.
\end{equation}
It is worth noting that, for the overall consistency of the likelihood definition, at least one MC event needs to be expected per bin, such that $N_{\text{bins}} \leq N_{\text{sim}}$, and since the system must have at least one $\theta$, $1 \leq N_{\text{alpha}}$.\\
To recap, event reweight factorization allows to reduce the number of intermediate calculations to update $\vec{\lambda}(\vec{\theta})$. In realistic HEP cases, $N_{\text{bins}}$ spans from 1 to several hundreds or thousands, $N_{\text{sim}}$ is often in the range of $10^5$--$10^8$ events, and $N_{\text{alpha}}$ can range from 1 to hundreds. When considering these numbers, it becomes clear that, systems allowing a high degree of factorization can benefit enormously from factorizing the calculation $\vec{\lambda}(\vec{\theta})$, dramatically reducing the number of necessary calculations.

%If a system can be fully factorized the number of calculations can be as low as $N^{\text{min}}_c$, and if it can not be factorized at all then one is forced to do $N^{\text{max}}_c$. In a generic case where only partial factorization is possible an nminimum number of calculation $N_c$ in between these two cases must be made. 

%not less than tens of thousands of events but more often it consists of hundreds of thousands or even millions of events, and $N_{\text{alpha}}$ can range from 1 to several hundreds. When considering these numbers, it becomes clear that, in systems allowing a high degree of factorization dramatic speed ups can be achieved.\\

\subsection*{Maximum factorization}
The goal is to derive a generic expression that allows to utilize the maximum possible amount of factorization in every system. Let's start by combining Eq.~\ref{eq:SmartEvtRwg} and Eq.~\ref{eq:RspFSmart}:

\begin{equation}
\label{eq:GeneralEq}
\lambda_i(\vec{\theta}) = \sum^{\text{i-events}}_k (w^{k}_{\text{nom}} \times  \prod_\alpha h^\alpha_{k}).
\end{equation}
Where for simplicity, $h^\alpha_{k}(\theta_\alpha)$ is shortened into $h^\alpha_{k}$. We have seen there are two possible types of events in each bin: those that share response functions with other events, and therefore can be factorized; and those that get unique event-by-event weights and can't be factorized. Let's make this explicit:
\begin{equation}
\label{eq:GeneralEqFactNoFact}
\lambda_i(\vec{\theta}) = \sum^{\text{i-fact}}_k (w^{k}_{\text{nom}} \times  \prod_\alpha h^\alpha_{k}) + \sum^{\text{i-no-fact}}_k (w^{k}_{\text{nom}} \times  \prod_\alpha h^\alpha_{k}).
\end{equation}
The second sum can not be simplified, so let's focus on the first. We note that:
\begin{equation}
\label{eq:ProdBreakUp}
     \prod_\alpha h^\alpha_{k} =
    \begin{cases}
      1 & \text{Trivial Case}\\
      h^\alpha & \text{Case A}\\
      h^\alpha h^\beta & \text{Case B}\\
      h^\alpha h^\beta h^\gamma & \text{Case C}\\
      \text{etc.} &
      % (\theta_1) \times h^{[2,3]}_{k}(\theta_2, \theta_3) & \text{case 1}\\
      % h^0(\theta_0) \times h^1 (\theta_1) &  \text{case 2}\\
      % h^0(\theta_0) \times h^{[2,3]}_{k}(\theta_2, \theta_3) & \text{case 3}\\
      % h^0(\theta_0) & \text{otherwise.}
    \end{cases}
\end{equation}
Where:
\begin{itemize}
    \item Trivial Case: Events that get no weights from any parameters.
    \item Case A: Events that get weights from a single $\alpha$.
    \item Case B: Events that get weights from two distinct parameter indices, $\alpha$ and $\beta$.
    \item Case C: Events that get weights from three distinct parameter indices, $\alpha$, $\beta$ and $\gamma$.
\end{itemize}
One can continue this list straightforwardly including the succeeding cases D, E, F, etc, if necessary. However, for the illustration here that is not necessary, and we limit ourselves to products involving up to three distinct parameter indices.\\
With the above considerations follows:
\begin{align}
\label{eq:GeneralEqFactTerm}
% \sum^{\text{i-fact}}_k (w^{k}_{\text{nom}} \times  \prod_\alpha h^\alpha_{k}) &= \sum^{\text{Trivial}}_k w^{k}_{\text{nom}} \nonumber \\
% &+\sum_{\{\alpha\}} \left(h^\alpha \times \sum^{\text{Case A$_{\{\alpha\}}$ }}_k w^{k}_{\text{nom}} \right) \nonumber \\
% &+\sum_{\{\alpha,\beta\}} \left(h^\alpha \times h^\beta \times \sum^{\text{Case B$_{\{\alpha,\beta\}}$ }}_k w^{k}_{\text{nom}} \right)  \nonumber \\
% &+\sum_{\{\alpha,\beta, \gamma\}} \left(h^\alpha \times h^\beta \times h^\gamma \times \sum^{\text{Case C$_{\{\alpha,\beta,\gamma\}}$    }}_k w^{k}_{\text{nom}} \right)  + \text{etc.}
\sum^{\text{i-fact}}_k (w^{k}_{\text{nom}} \times  \prod_\alpha h^\alpha_{k}) &= \sum^{\text{Trivial}}_k w^{k}_{\text{nom}} \nonumber \\
&+\sum_{\{\alpha\}} \left(h^\alpha \sum^{\text{Case A$_{\{\alpha\}}$ }}_k w^{k}_{\text{nom}} \right) \nonumber \\
&+\sum_{\{\alpha,\beta\}} \left(h^\alpha h^\beta \sum^{\text{Case B$_{\{\alpha,\beta\}}$ }}_k w^{k}_{\text{nom}} \right)  \nonumber \\
&+\sum_{\{\alpha,\beta, \gamma\}} \left(h^\alpha h^\beta h^\gamma \sum^{\text{Case C$_{\{\alpha,\beta,\gamma\}}$    }}_k w^{k}_{\text{nom}} \right)  + \text{etc.}
\end{align}
Where curly brackets indicate unique sets without permutations of parameter indices. If we multiply and divide by $\lambda_i(\vec{\theta}_{\text{nom}})$, follows:
\begin{align}
\label{eq:CompactGeneralEqFactTerm}
&\sum^{\text{i-fact}}_k (w^{k}_{\text{nom}} \times \prod_\alpha h^\alpha_{k}) =  \nonumber \\
&= \lambda_i(\vec{\theta}_{\text{nom}}) \left(C_T +\sum_{\{\alpha\}} h^\alpha C_\alpha +\sum_{\{\alpha\beta\}} h^\alpha h^\beta C_{\alpha\beta} +\sum_{\{\alpha\beta\}} h^\alpha h^\beta h^\gamma C_{\alpha\beta\gamma} + \text{etc.} \right)
\end{align}
Where, for the $i$-th bin, $C_T$ denotes the fraction of nominal events that are trivial, each C$_\alpha$ denotes the fraction of nominal events that only get weights from one specific $\alpha$, C$_{\alpha\beta}$ denotes the fraction of nominal events that only get weights from one specific unique combination of $\alpha$ and $\beta$, etc.\\
Finally, let's plug Eq.~\ref{eq:CompactGeneralEqFactTerm} in Eq.~\ref{eq:GeneralEqFactNoFact} to provide the final generic expression:
\begin{align}
\label{eq:FinalEq}
\lambda_i(\vec{\theta}) &= \sum^{\text{i-no-fact}}_k (w^{k}_{\text{nom}} \times  \prod_\alpha h^\alpha_{k}) + \nonumber \\
&+ \lambda_i(\vec{\theta}_{\text{nom}}) \left(C_T +\sum_{\{\alpha\}} h^\alpha C_\alpha +\sum_{\{\alpha\beta\}} h^\alpha h^\beta C_{\alpha\beta} +\sum_{\{\alpha\beta\}} h^\alpha h^\beta h^\gamma C_{\alpha\beta\gamma} + \text{etc.} \right)
\end{align}
For any system, Eq.~\ref{eq:CompactGeneralEqFactTerm} maximally factorizes the expression of the reweighted event rate prediction in every bin, minimizing the number of calculations necessary to evaluate the likelihood.

\section{A toy problem}
For illustration, let's consider the following toy problem. Paired with the explanation, I provide a publicly available implementation\footnote{\url{https://github.com/cesarjesusvalls/factorization_demo}} meant to demonstrate explicitly how to put the practice the concepts discussed in this article.\\
Imagine that we have an accelerator neutrino experiment, that consists of only one Cherenkov detector and a neutrino beam. Consider the beam to be identical to that of T2K~\cite{T2K:2011qtm} with our hypothetical Cherenkov detector placed at a distance short enough as to safely neglect neutrino oscillations. Then, we simulate MC samples using GENIE~\cite{Andreopoulos:2009rq} together with a public T2K flux release~\cite{T2K:2023smv}. Although this is by no means necessary for this toy problem, to mimic a pseudo-realistic physics case, the output true muon angle and momentum from GENIE is smeared by 10\% and a simplified 1 muon ring selection is applied using the following criteria: 1) the reconstructed muon momentum $p^{\text{reco}}_\mu$ satisfies  $0.25 \leq p_\mu \leq 2$GeV/c 2) there are no $\pi^0$ in the final state abd 3) there are no charged pions with momentum above 250 MeV/c.\\
\begin{figure}[ht!]
\centering
\centering\includegraphics[width=0.6\textwidth]{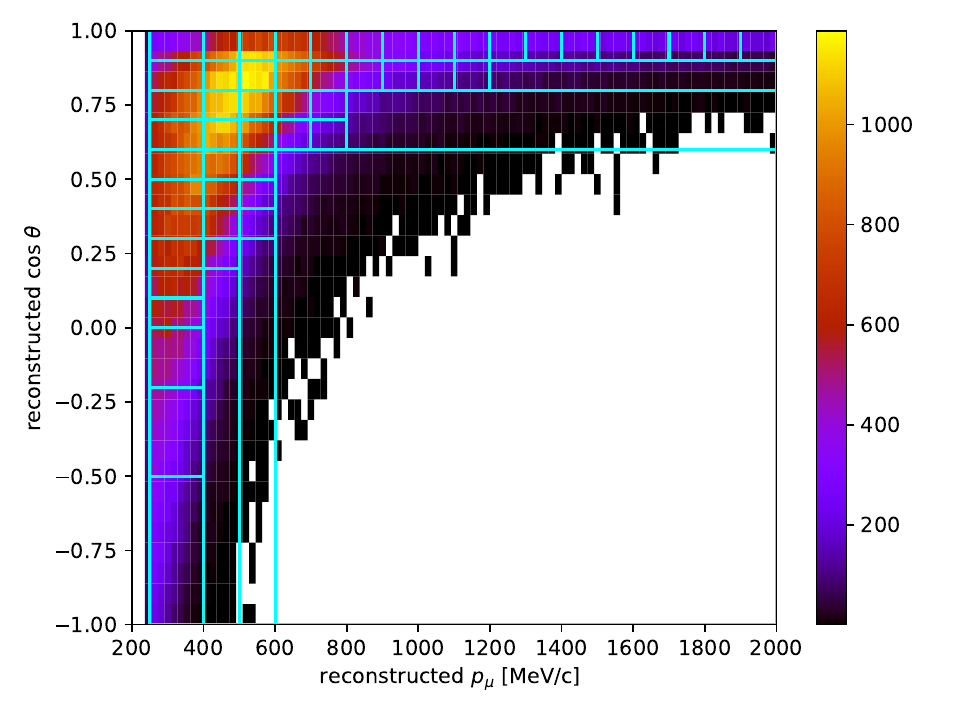}
\caption{Illustration of the toy problem sample described in the text. The cyan lines illustrate the edges of its 57 bins.}
\label{fig:distribution}
\end{figure}
We bin the reconstructed muon angle and momentum distributions in fifty seven 2D regions. The binning, presented in Fig.~\ref{fig:distribution}, is chosen as to not be far off from a realistic choice for a real experiment. Nevertheless, in the context of this toy problem the binning choice can be considered as arbitrary. We then consider several choices for $\vec{\theta}$:
\begin{itemize}
    \item \texttt{Test\_0}: Only one normalization parameter modifies charged-current quasielastic (CCQE) events get.
    \item \texttt{Test\_1}: For parameters, divided in three regions in true neutrino energy are used, to mimic flux reweighting.
    \item \texttt{Test\_2}: Corresponds to all parameters (4 in total) from \texttt{Test\_0} and \texttt{Test\_1}.
    \item \texttt{Test\_2}: An intermediate number of parameters (12 in total): Two cross section normalization CCQE and 2p2h channels, and 10 flux regions. 
    \item \texttt{Test\_3}: A large number of parameters (51 in total): All cross section modes are considered, there are 11 in the sample under study, and 40 flux regions.
\end{itemize}
After successfully verifying that both schemes --w/ and w/o factorization-- lead to identical binned event rates for any $\vec{\theta}$, a series of speed test were done. The results are summarizes in Table~\ref{tab:speed}. 

\begin{table}[ht!]
    \centering
    \begin{tabular}{ |l||r|r|r|r|r|r| }
     \hline
     \hline
     Test ID & U. Conf & $\sum$ U. Conf. & W. Calc. & T. w/ Fact & T. w/o Fact & Speed Up \\
     \hline
     % 0 & 2   &   114   & 78$\pm$18  \textmu s & 259$\pm$3 ms   & $\times$3301\\
     % 1 & 3   &   162   & 107$\pm$20 \textmu s & 284$\pm$4 ms   & $\times$2659\\
     % 2 & 6   &   278   & 163$\pm$20 \textmu s & 320$\pm$6 ms   & $\times$1783\\
     % 3 & 29  &  1027   & 473$\pm$87 \textmu s & 322$\pm$3 ms   & $\times$681\\
     % 4 & 308 &  6394   & 3038$\pm$139 \textmu s & 338$\pm$6 ms & $\times$113\\
    0 	& 2 	& 114 	& 200992 	& 115 $\pm$ 13 \textmu s 	& 349 $\pm$ 5 ms 	& $\times$3044 \\
    1 	& 3 	& 162 	& 285460 	& 165 $\pm$ 18 \textmu s 	& 405 $\pm$ 5 ms 	& $\times$2454 \\
    2 	& 6 	& 278 	& 486452 	& 263 $\pm$ 23 \textmu s 	& 451 $\pm$ 10 ms 	& $\times$1713 \\
    3 	& 29 	& 1027 	& 525824 	& 810 $\pm$ 41 \textmu s 	& 460 $\pm$ 7 ms 	& $\times$568 \\
    4 	& 308 	& 6394 	& 570920 	& 5411$\pm$ 234 \textmu s 	& 474 $\pm$ 10 ms 	& $\times$88 \\
     \hline
     \hline
    \end{tabular}
    \caption{Summary metrics for the different tests in the toy problem. From left to right: The number of unique parameter configurations (U. Conf), the sum of the number of unique parameter configurations in each bin ($\sum$ U. Conf.), the number of weight calculations without using factorization (W. Calc.), the time to calculate $\vec{\lambda}(\vec{\theta})$ with (T. w/ Fact) and without (T. w/o Fact) factorization, and the speed up improvement calculated as the ratio of the last two columns.}
    \label{tab:speed}
\end{table}

\begin{figure}[ht!]
\centering
\centering\includegraphics[width=0.49\textwidth]{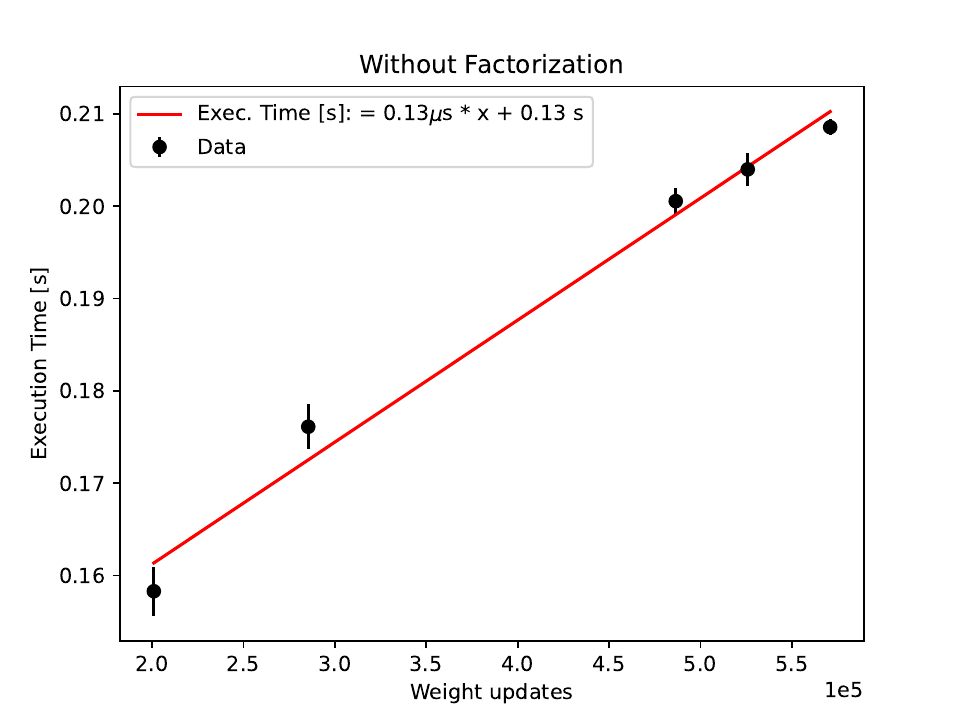}
\centering\includegraphics[width=0.49\textwidth]{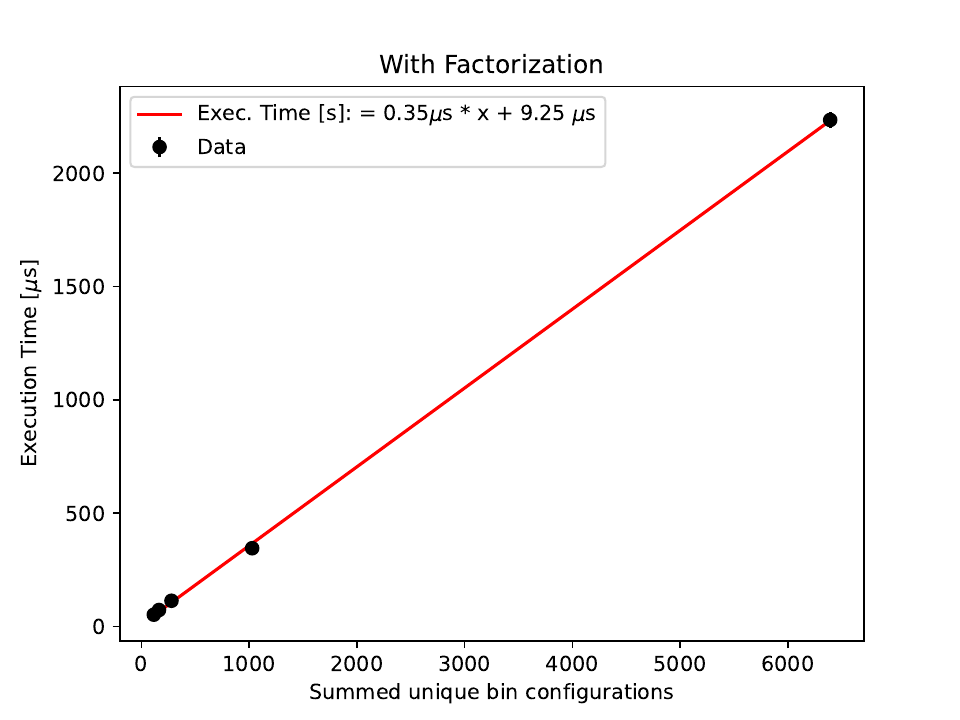}
\caption{Execution times associated to the tests in Table~\ref{tab:speed} for event reweights without (left) and with (right) the application of factorization in the calculation.}
\label{fig:time_response}
\end{figure}

\noindent As expected, the reweight time without factorization, remains on the same order of magnitude for all tests as the leading contributor to the execution time are operations that are performed once for all events, such as searching the parameter indices associated to each event and updating the event weight. There is a linear sub-leading time dependence with the number of response function evaluations, necessary to calculate the weight updates, which translates into a slightly increasing the execution time from \texttt{Test\_0} to \texttt{Test\_4}, as observed in the left panel of Fig.~\ref{fig:time_response}. In contrast, the execution time with factorization depends directly on the number of bins and the number of unique configurations in each bin. The execution time grows linearly with the sum of the unique parameter configurations in all bins, as illustrated in the right panel of Fig.~\ref{fig:time_response}. This markedly different computational behavior results in drastically improved execution times for all tests, illustrating the importance of using event factorization.

\section{Conclusions}
In this article the importance of factorizing unique parameter configurations has been explained and a general formula for its application in any system involving event reweight has been presented. For illustration, a publicly available toy model has been prepared and used to report in the article various metrics that exemplify the advantages of incorporating factorization in the implementation of likelihood-based analysis software. Such statistical analyses are very common in HEP, where the computation time often slows down measurements within large experimental collaborations. Therefore, this article aims to guide readers on effectively utilizing factorization in calculations related to event reweighting, thereby accelerating studies and reducing its associated computational carbon footprint.

\section{Acknowledgments}
The author acknowledges fruitful discussions with A. Blanchet, L. Berns and A. Muñoz.

\pagebreak

\bibliographystyle{apsrev4-2}
\bibliography{bibliog}

\end{document}